\documentstyle[12pt]{article}

\catcode`\@=11
\long\def\@makefntext#1{
\protect\noindent \hbox to 3.2pt {\hskip-.9pt
$^{{\ninerm\@thefnmark}}$\hfil}#1\hfill}		

 \def\@makefnmark{\hbox to 0pt{$^{\@thefnmark}$\hss}}  

\def\ps@myheadings{\let\@mkboth\@gobbletwo
\def\@oddhead{\hbox{}
\rightmark\hfil\ninerm\thepage}
\def\@oddfoot{}\def\@evenhead{\ninerm\thepage\hfil
\leftmark\hbox{}}\def\@evenfoot{}
\def\sectionmark##1{}\def\subsectionmark##1{}}

\renewcommand{\section}[1] {\vspace{0.6cm}\addtocounter{sectionc}{1}
\setcounter{subsectionc}{0}\setcounter{subsubsectionc}{0}\noindent
	{\bf\thesectionc. #1}\par\vspace{0.4cm}}
\renewcommand{\subsection}[1] {\vspace{0.6cm}\addtocounter{subsectionc}{1}
	\setcounter{subsubsectionc}{0}\noindent
	{\it\thesectionc.\thesubsectionc. #1}\par\vspace{0.4cm}}
\renewcommand{\subsubsection}[1]
{\vspace{0.6cm}\addtocounter{subsubsectionc}{1}
	\noindent {\rm\thesectionc.\thesubsectionc.\thesubsubsectionc.
	#1}\par\vspace{0.4cm}}

\newcounter{appendixc}
\newcounter{subappendixc}[appendixc]
\newcounter{subsubappendixc}[subappendixc]

\renewcommand{\appendix}[1] {\vspace{0.6cm}
        \refstepcounter{appendixc}
        \setcounter{figure}{0}
        \setcounter{table}{0}
        \setcounter{equation}{0}
        \renewcommand{\thefigure}{\Alph{appendixc}.\arabic{figure}}
        \renewcommand{\thetable}{\Alph{appendixc}.\arabic{table}}
        \renewcommand{\theappendixc}{\Alph{appendixc}}
        \renewcommand{\theequation}{\Alph{appendixc}.\arabic{equation}}
        \noindent{\bf Appendix \theappendixc #1}\par\vspace{0.4cm}}

\def\abstracts#1{{
	\centering{\begin{minipage}{30pc}\tenrm\baselineskip=12pt\noindent
	\centerline{\tenrm ABSTRACT}\vspace{0.3cm}
	\parindent=0pt #1
	\end{minipage}}\par}}

\renewenvironment{thebibliography}[1]
	{\begin{list}{\arabic{enumi}.}
	{\usecounter{enumi}\setlength{\parsep}{0pt}
\setlength{\leftmargin 1.25cm}{\rightmargin 0pt}
	 \setlength{\itemsep}{0pt} \settowidth
	{\labelwidth}{#1.}\sloppy}}{\end{list}}

\newcounter{itemlistc}
\newcounter{romanlistc}
\newcounter{alphlistc}
\newcounter{arabiclistc}

\newcommand{\fcaption}[1]{
        \refstepcounter{figure}
        \setbox\@tempboxa = \hbox{\tenrm Fig.~\thefigure. #1}
        \ifdim \wd\@tempboxa > 6in
           {\begin{center}
        \parbox{6in}{\tenrm\baselineskip=12pt Fig.~\thefigure. #1}
            \end{center}}
        \else
             {\begin{center}
             {\tenrm Fig.~\thefigure. #1}
              \end{center}}
        \fi}

\newcommand{\tcaption}[1]{
        \refstepcounter{table}
        \setbox\@tempboxa = \hbox{\tenrm Table~\thetable. #1}
        \ifdim \wd\@tempboxa > 6in
           {\begin{center}
        \parbox{6in}{\tenrm\baselineskip=12pt Table~\thetable. #1}
            \end{center}}
        \else
             {\begin{center}
             {\tenrm Table~\thetable. #1}
              \end{center}}
        \fi}

\def\@citex[#1]#2{\if@filesw\immediate\write\@auxout
	{\string\citation{#2}}\fi
\def\@citea{}\@cite{\@for\@citeb:=#2\do
	{\@citea\def\@citea{,}\@ifundefined
	{b@\@citeb}{{\bf ?}\@warning
	{Citation `\@citeb' on page \thepage \space undefined}}
	{\csname b@\@citeb\endcsname}}}{#1}}

\newif\if@cghi
\def\cite{\@cghitrue\@ifnextchar [{\@tempswatrue
	\@citex}{\@tempswafalse\@citex[]}}
\def\citelow{\@cghifalse\@ifnextchar [{\@tempswatrue
	\@citex}{\@tempswafalse\@citex[]}}
\def\@cite#1#2{{$\null^{#1}$\if@tempswa\typeout
	{IJCGA warning: optional citation argument
	ignored: `#2'} \fi}}

\def\fnt#1#2{\footnotetext{\kern-.3em
	{$^{\mbox{\sevenrm #1}}$}{#2}}}

 1
 1
 1

\font\tenbf=cmbx10
\font\tenrm=cmr10
\font\tenit=cmti10

\font\ninerm=cmr9

\begin{document}

\centerline{\tenbf ARE THERE SOLITONS IN THE TWO-HIGGS
STANDARD MODEL? \footnote{talk at the joint US-Polish workshop
on {\it Physics from the Planck scale to the electroweak scale},
Warsaw, September 1994.}
}
\baselineskip=16pt
\centerline{\tenbf  }
\centerline{\ninerm  }
\vspace{0.8cm}
\centerline{\tenrm C.P. BACHAS}
\baselineskip=13pt
\centerline{\tenit Centre de Physique Th\'eorique}
\baselineskip=12pt
\centerline{\tenit Ecole Polytechnique}
\baselineskip=12pt
\centerline{\tenit 91128 Palaiseau, France}
\vspace{0.3cm}
\centerline{\tenrm and}
\vspace{0.3cm}
\centerline{\tenrm T.N. TOMARAS}
\baselineskip=13pt
\centerline{\tenit Physics Department, University of Crete}
\baselineskip=12pt
\centerline{\tenit and Research Center of Crete}
\baselineskip=12pt
\centerline{\tenit 714 09 Heraklion, Greece}
\vspace{0.9cm}
\abstracts{ We present some evidence, based on the analysis
of lower-dimensional models,
 for the possible existence of classically-stable
winding solitons in the two-higgs electroweak theory.}

\vfil
\vskip 0.8cm
\rm\baselineskip=14pt

The search for stable lumps in the Weinberg-Salam model
has a long history.
It has revealed a rich
structure of classical solutions including the sphaleron
\cite{DHN,Taubes,Klink,Boguta} ,
deformed sphalerons \cite{eila,Brihaye,Yaffe,Klink2}
 and vortex strings \cite{Nambu,Soni,Vach,Vach1,Periv}.
Such solutions could play a role in understanding
{\ninerm (B+L)}-violation
and structure formation in the early universe,
but they are all classically-unstable or/and extended.
They have therefore
no direct  present-day manifestation,
contrary to long-lived particles whose
relic density could at least in principle be
 detected.

The existence of  particle-like excitations has, on the
other hand, been argued for in the context of a strongly-
interacting higgs sector \cite{Tze,Fahri,Rub,Zee,Carlson}.
The advocated particles can be thought of as
technibaryons of
an underlying technicolor model.
They are described in
bosonic language by winding solitons of
 an effective non-renormalizable
lagrangian for the   pseudo-goldstone-boson
 (or technipion) field,
much like  skyrmions\cite{Skyrme}
 of the effective chiral lagrangian
of $QCD$.  This is of course a phenomenological
 description, since
the   properties of such hypothetical
particles cannot be
calculated reliably within a semi-classical expansion.
Furthermore, in
view of the difficulties
facing technicolor models,
the  possibility of a strongly-interacting higgs sector is
not theoretically  appealing.

It would be clearly    more interesting
 if classically-stable winding excitations
could  arise in a {\it weakly-coupled} scalar
 sector.
To be more precise let us decompose the
 higgs-doublet  field into a
real (positive)  magnitude   and a
group-phase: $\Phi = F U$, and
consider static configurations with $U(x)$
 wrapping $N$ times
around the $SU(2)$ manifold. These are
 potentially unstable
for at least three distinct reasons:
\break
{\it (a)} because $N$ is not   conserved
whenever the magnitude $F$   goes through
zero;\break
{\it (b)} because $N$ is not gauge-invariant and can,
in particular, be non-vanishing
even in a vacuum state;
and {\it (c)} because scalar-field configurations can loose
their energy
by shrinking to zero size \cite{Derrick}.
We refer to these for short as the {\it radial, gauge}
 and {\it scale}
instabilities. They can be eliminated formally by
{\it (a)}
 taking the physical-higgs mass $m_H\to\infty$,
{\it (b)} decoupling the electroweak gauge fields,
 and
{\it (c)} adding appropriate higher-derivative terms to the
 action.
The question is whether classical stability can
 be maintained
while relaxing the above conditions.
This
 has been investigated numerically in the
{\it minimal} case of one
doublet: although one may indeed relax both
the weak gauge coupling \cite{Rub,eila}
and  the higgs mass \cite{Kunz}
  up to some finite critical values,
  stability cannot apparently be achieved without the
   non-renormalizable  higher-derivative
  terms in the
 action.
On the other hand, as we have
 demonstrated recently,
  metastable winding solitons do arise in
renormalizable models in
two \cite{mexican} and three  \cite{preprint}
space-time dimensions. The way this happens
  is we believe instructive and could guide
the search for such
semi-classical solitons in four dimensions.

The simplest context in which the {\it radial}
instability is an issue is a two-dimensional model of
  a complex-scalar field with mexican-hat potential:
$ V = {1\over 4}\lambda (\Phi^*\Phi - v^2)^2$.
To find winding solitons
we must take space to be periodic with period
$L$ \footnote{Alternatively we may add a mass term:
$\delta V = -\mu^2 v Re(\Phi)$,  that
lifts the vacuum degeneracy. Stable winding
 excitations, which reduce to
the sine-Gordon solitons in the $\lambda\to\infty$ limit,
can be shown \cite{Pallis} numerically to
 exist for $\lambda v^2/\mu > 18.8$ \  .}
{}.
 The condition for
  classical stability can in this case
be derived analytically  and reads\cite{mexican}:
$ m_H L > \sqrt{5}  \ ,  $
where $m_H=\sqrt{2\lambda} v$. The
 classically-relevant parameter
is thus the radial-higgs mass in units of the soliton  size.
This follows also   by
comparing  the loss in potential energy to the gain
in gradient energy when trying to undo the winding by
reducing  the magnitude of the scalar. Note that
the loop-expansion parameter $\lambda L^2$,
  can be taken to zero independently so as to reach a
semiclassical limit.

 The above winding solitons become unstable
  classically if we gauge the $U(1)$ symmetry of the model.
The {\it gauge} instability is in fact
more severe than in four dimensions,
because   no energetic barrier
  opposes the turning-on of
a static space-like gauge field, which is necessary
 to reach a winding-vacuum
state. The minimal abelian-higgs model has thus only unstable
(sphaleron) solutions \cite{Giller}
 \footnote{It was claimed erroneously in [22] that it
has no static solutions whatsoever.
 This is only correct in the $\lambda\to\infty$
limit.}$\ $ .
The situation changes, however, drastically
if there are more than one complex scalars. The
gauge-invariant  relative phases
of any two of them   cannot in this case  wind around
non-trivially in a vacuum state.
 An explicit
analysis
of this extended abelian-higgs model \cite{mexican}
 shows that   winding solitons
  persist down to scalar masses close to
 the inverse soliton size: gauging and the extra higgs
  enhance the stability region found in the global model.

 The {\it scale} instability becomes an issue for
 the first time
in three space-time dimensions.
To be more precise we consider a real-triplet
 scalar field $\Phi_a(x)$
($a=1,2,3$) with mexican-hat
potential : $V= {1\over 4}\lambda (\Phi_a\Phi_a - v^2)^2$.
The  limit $\lambda\to\infty$ corresponds to the
$O(3)$
non-linear $\sigma$-model. This is known to possess
 winding solitons, characterized by non-trivial
 mappings of the two-sphere
onto itself, and having
  arbitrary size \cite{Belavin}.
For finite $\lambda$ on the other hand, or in the
 presence of a
symmetry-breaking potential, Derrick's
 scaling argument \cite{Derrick}
 shows that these   solitons are unstable to shrinking.
One can of course again invoke higher-derivative terms
to stabilize the scale\cite{Zak}.
The same result is however in this case  achieved by a massive
$U(1)$ gauge field with only renormalizable
 couplings \cite{preprint}.
This can be established
 by perturbing around the
$O(3)$ non-linear $\sigma$-model limit,
 or else by solving numerically
the equations of motion.

What do these lower-dimensional solitons teach us?
First,
they are interesting in their own right,
 since they correspond
to a new class of wall  and string defects
 in renormalizable
four-dimensional models.
Second,  they suggest by analogy
 that classically-stable winding solitons may
  exist in a weakly-coupled two-higgs extension
 of the standard model.
These hypothetical
solitons would: {\it (b)} be characterized
 by the non-trivial winding
of the relative phase of the two doublets, and thus be immune to
the gauge mode of decay;
{\it (c)} have a scale stabilized by
  electroweak
magnetic fields and hence  of order $1/m_W$;
and {\it (a)}   hopefully stay stable for higgs
 masses near $m_W$
and thus compatible with perturbative unitarity
 \footnote{Though
 admittedly premature, some other
  physical properties of such would-be particles are fun to
contemplate: being classically stable they could easily have
cosmological life times. They would have a
 mass in the $\sim 10\  TeV$
region, zero charge and dipole moments in their ground state,
and geometrical interaction cross sections of order $1/m_W^2$.
Assuming maximum production at the
 electroweak phase transition,
a rough estimate of their present abundance shows that they
could be candidates for cold dark matter in the
universe.}$\ $.
Mathematically the situation is the same as in the
hidden-gauge-boson models \cite{Kunzz,Dobado} of strong
 and electroweak
 interactions,
except that the role of the hidden gauge bosons is here played
by  $W^{\pm}$ and $Z$ themselves.  Although
previous numerical investigations \cite{Kunzz,Peccei}
have shown no sign of stable solitons in these contexts,
  a systematic search is in our opinion necessary in order to
settle definitely the issue \cite{Tinyakov}.
\vskip 0.3cm

{\bf Acknowledgements}
\vskip 0.2cm

This research was supported in part by
 the EEC grants CHRX-CT94-0621 and CHRX-CT93-0340,
as well as by the Greek General Secretariat
 of Research and Technology
grant 91$E\Delta$358.

\end{document}